# The KV Cache Working Set: Online Capacity Planning for LLM Inference Systems

Luchang Li*, Shuaishuai Wang, Zhao Ruan, Dongfang Li, Bozhao Gong
Kingsoft Cloud

**Abstract**

Prefix caching is critical for efficient large language model (LLM) serving, particularly for agentic workloads that repeatedly invoke the model with a growing conversation and tool-use history. By reusing the key-value (KV) states of previously processed prefixes, prefix caching avoids redundant prefill computation. Its effectiveness, however, depends on retaining a sufficiently large set of KV cache states. Provisioning enough cache to preserve all historical KV states is prohibitively expensive and often unnecessary, whereas insufficient capacity can substantially degrade the cache hit rate. Determining the KV cache working set, defined as the minimum cache capacity required to achieve a target hit rate, is therefore essential for efficient cache provisioning and system design.

We present KVSET, an online analyzer that estimates the KV cache working set of LLM serving workloads. KVSET uses the Mattson stack algorithm to efficiently estimate cache hit rates across a wide range of cache capacities. For each KV cache page, KVSET computes its LRU stack distance and compares it with the page number of each candidate capacity. This comparison determines whether the page would be a hit at each capacity without independently simulating every capacity configuration. KVSET therefore substantially reduces the computational and memory overhead of conventional capacity-by-capacity simulation and makes online working-set analysis practical.

KVSET further determines the minimum cache capacity based on the maximum LRU depth among the prefix pages required to achieve the target hit rate. We validate KVSET using traces collected from production LLM workloads and show that its estimates closely match measurements from real cache deployments. The open-source implementation supports both online request processing and offline trace replay, enabling practical KV cache analysis and provisioning for LLM serving systems.

## 1 Introduction

Advances in large language model (LLM) reasoning, planning, tool use, and multi-turn interaction have made LLM-based agents a common paradigm for applications such as coding assistants, deep research, customer service[5, 11, 14, 16, 18]. Completing a single agentic task often requires tens or hundreds of interaction turns, and each new turn typically reuses most or all of the input and output history from preceding turns[3, 9, 12, 13]. This repeated prefix reuse makes prefix caching particularly effective: reusing previously computed key-value (KV) states avoids redundant prefill computation, improves serving throughput, and reduces inference cost. The throughput benefit is nonlinear: as shown in Figure 1, prefill throughput rises sharply at high KV cache hit rates rather than scaling linearly with the hit rate. Ignoring memory-access overhead,

* Corresponding author: liluchang@kingsoft.com

Equation 1 gives the relationship between prefill throughput and cache hit rate, where $T$ is the prefill throughput at hit rate $r$, and $\widehat{T}$ is the prefill throughput with no KV cache hits. A high KV cache hit rate is therefore critical to LLM serving performance and cost efficiency. Modern inference engines, including vLLM[6] and SGLang[17], provide built-in KV cache management, while systems such as Mooncake[10] and LMCache[7] further extend KV cache capacity by supporting multiple storage tiers, heterogeneous storage media, and global cache sharing.

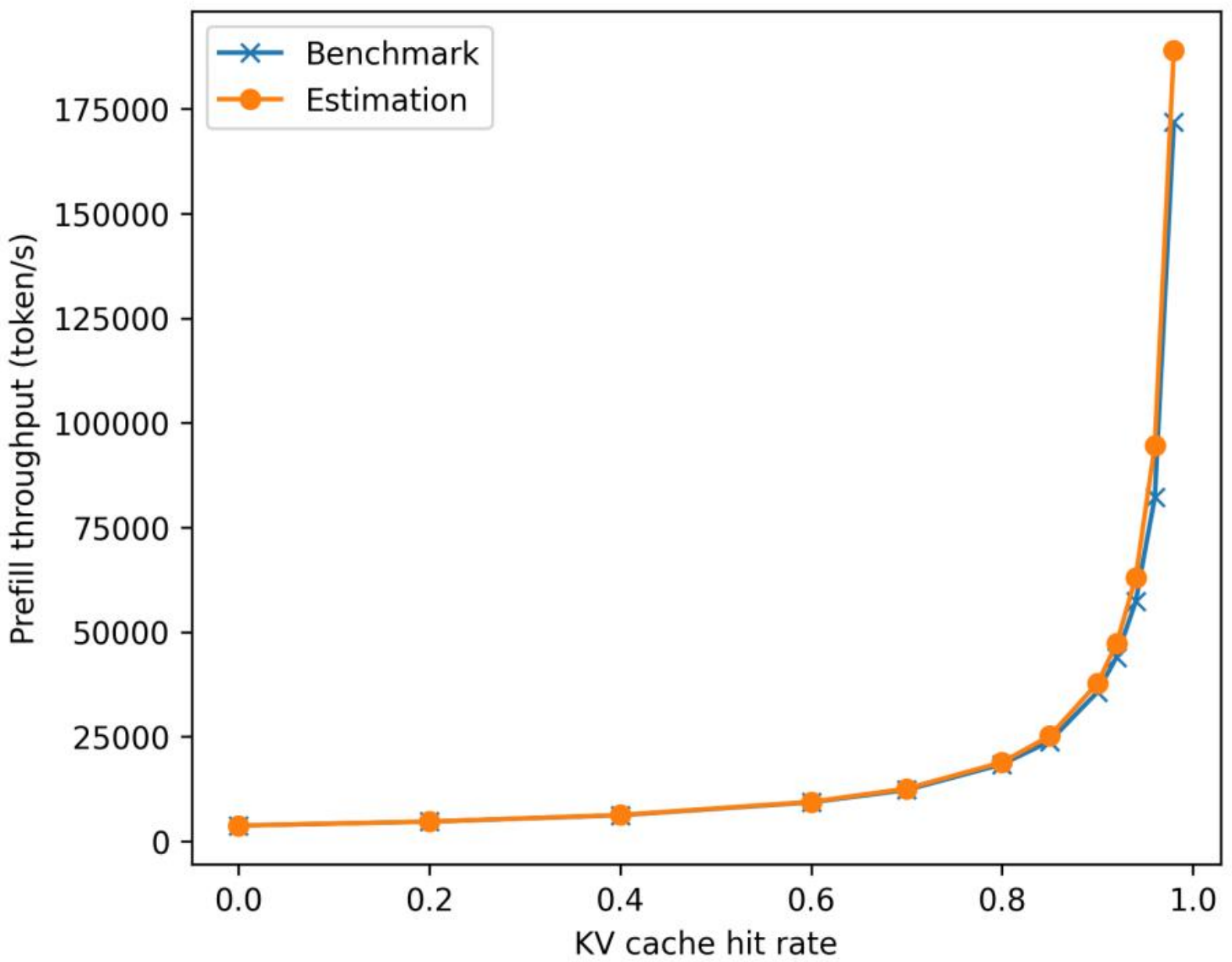

Figure 1. Prefill throughput as a function of KV cache hit rate. Benchmark use an NVIDIA H20 GPU server, the GLM-5.2 model with W4A8 quantization, and a 16K-token input. Estimated throughput is computed from the zero-hit throughput using Equation 1.

$$\mathrm{T} = \frac{\widehat{\mathrm{T}}}{1-\mathrm{r}} \tag{1}$$

High hit rates, however, are not free. Every retained token consumes KV storage across all transformer layers, and the footprint can reach dozens of KiB per token for a large model. Keeping the KV states of every historical request causes storage consumption to grow with trace length and may require an impractically large DRAM or SSD pool. Such provisioning is usually unnecessary because future hits depend primarily on a smaller hot working set, while pages outside that set may never be reused and can therefore be evicted.

Determining the KV cache capacity required by a workload both efficiently and accurately remains challenging. Measuring the required capacity by deploying physical cache pools at multiple sizes is prohibitively expensive. A more practical alternative is to simulate incoming requests online. Existing approaches, such as kvcache-simulator[19], estimate the capacity needed to achieve a target hit rate by independently replaying the workload against multiple simulated KV cache pools of different sizes. However, obtaining a sufficiently precise capacity estimate typically requires simulating dozens of candidate cache sizes. This repeated replay imposes substantial compute and latency overhead, making real-time integration with online serving systems difficult.

We address these challenges with KVSET, an efficient capacity-analysis framework for page-granular LRU prefix caches. LRU is one of the most widely adopted eviction policies in practical KV cache storage systems. Under LRU, the Mattson stack algorithm[1, 8] determines whether a page is resident at any cache capacity directly from its stack distance. This allows KVSET to evaluate many candidate capacities simultaneously without independently replaying the eviction behavior of a separate cache pool for each capacity. To reduce analysis overhead further, KVSET replaces the explicit LRU stack with a Fenwick tree[2], which computes stack distances efficiently. As a result, KVSET avoids maintaining and simulating multiple cache instances while still providing exact hit-rate estimates for a wide range of cache capacities, making real-time analysis of production request streams practical.

This paper makes the following contributions:

**Efficient KV cache capacity analysis.** We develop an efficient capacity-analysis method based on the Mattson stack algorithm. In a single pass over each request, our method derives hit rates for multiple KV cache capacities and identify the minimum capacity that achieves a target hit rate. This eliminates the need for simulating each cache capacity independently.

**Accurate online capacity estimation.** Experiments show that KVSET accurately predicts the KV cache capacity required to achieve a target hit rate. Its low computational and memory overhead also enables real-time analysis of online request streams.

**Open-source implementation.** We provide an open-source implementation that supports online request processing and offline trace replay. To the best of our knowledge, this is the first open-source tool that enables online KV cache capacity analysis. The code is available at https://github.com/llc-kc/kv_cache_capacity_estimator.

## 2 Method

Modern inference and KV cache systems, including vLLM and Mooncake, commonly use least recently used (LRU) eviction to manage limited KV cache capacity. Under LRU, cached pages are logically ordered by recency of access: the most recently accessed page resides at the top of the LRU stack, while pages nearer the bottom have been accessed less recently. Upon each access, the corresponding page is moved to the top of the stack. The Mattson stack algorithm exploits this stack property of LRU to determine whether an access would result in a cache hit under an arbitrary cache capacity. Specifically, a page is retained in a cache of capacity C if its LRU stack distance is smaller than C; otherwise, it would have been evicted before the current access.

As illustrated in Figure 2, KVSET uses the Mattson stack algorithm to simultaneously characterize KV cache eviction behavior across different cache capacities. To compute each page's LRU stack distance efficiently, KVSET records the page's most recent access position and maintains access positions in a Fenwick tree. For each access, the tree counts the distinct pages accessed more recently than the referenced page, which yields its LRU stack distance.

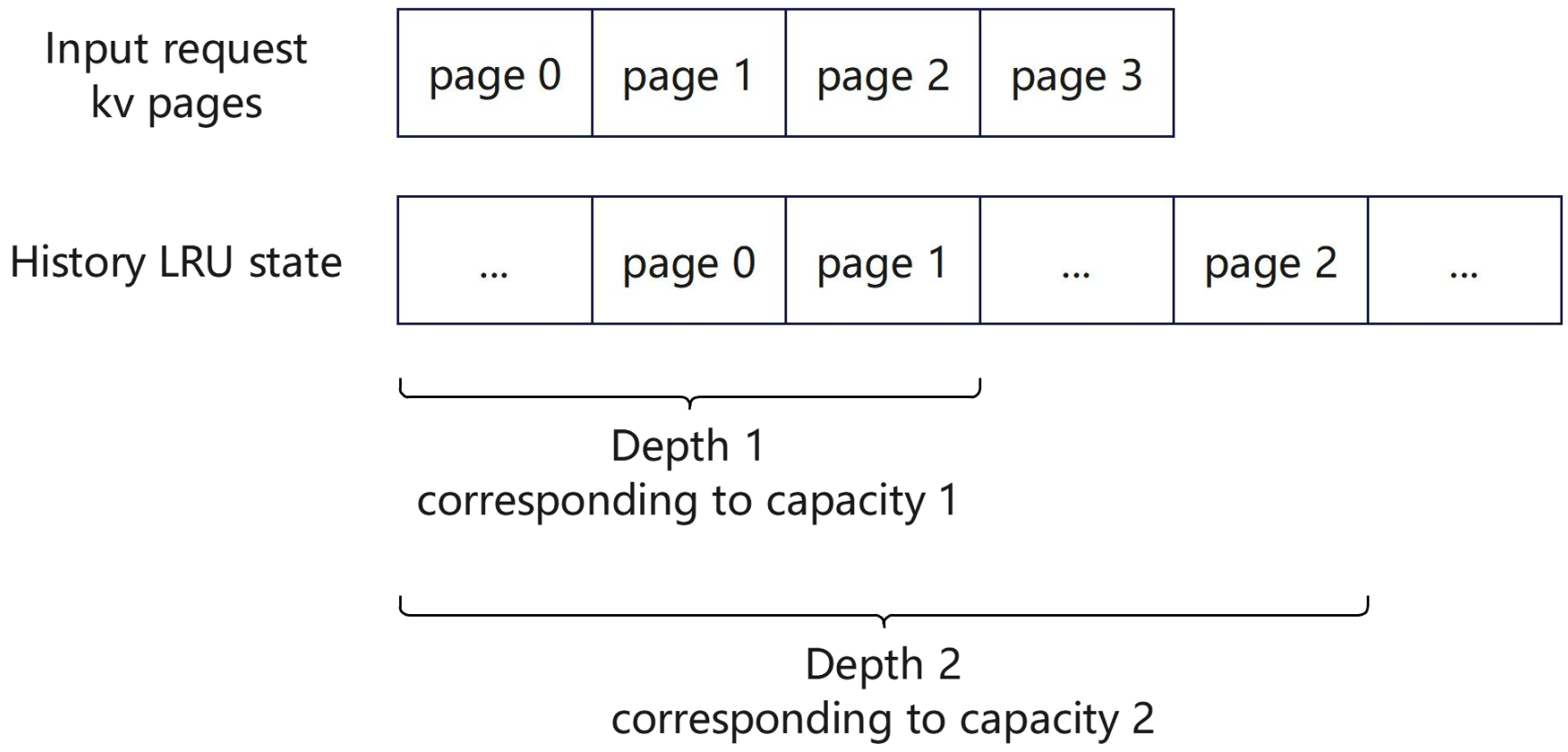

Figure 2. Using the Mattson stack algorithm to determine request-level KV cache hit rates at different capacities. For the illustrated request, pages 0 and 1 hit with a capacity 1, whereas page 2 hits with a capacity of 2 but is evicted with a capacity 1.

Given the stack distances of all KV pages in a request, KVSET can determine the hit or miss outcome of each access under any cache capacity without independently replaying the request for each capacity. Aggregating these results yields the entire capacity-to-hit-rate curve in a single replay. Furthermore, based on the maximum LRU depth among the prefix pages required to achieve the target hit rate, KVSET identify the minimum KV cache capacity required for each request. For example, in Figure 2, if the request is expected to hit two pages, Capacity 1 is sufficient; if three pages are expected to hit, the cache must be enlarged to Capacity 2.

# 3 Evaluation

**Capacity Estimation Accuracy**

We evaluate the accuracy of KVSET's hit-rate estimates across cache capacities. Our evaluation uses a trace of 10,000 requests collected from an internal production coding-agent workload. We use KVSET to predict the hit rate at each capacity and validate the predictions against measurements from deployed caches. Specifically, we deploy Mooncake and configure KV cache pools of multiple physical capacities for SGLang infernce engine. The experiments are conducted on NVIDIA H20 GPUs using GLM-5.2 with W4A8 quantization. MTP is disabled, and the KV cache is quantized to FP8, resulting in approximately 60 KB of KV cache storage per token in SGLang.

Figure 3 compares KVSET's predicted hit rates with measurements from the deployed caches. The predictions closely match the measurements across capacities, showing that KVSET accurately estimates the capacity-to-hit-rate relationship without explicitly deploying or simulating multiple cache configurations.

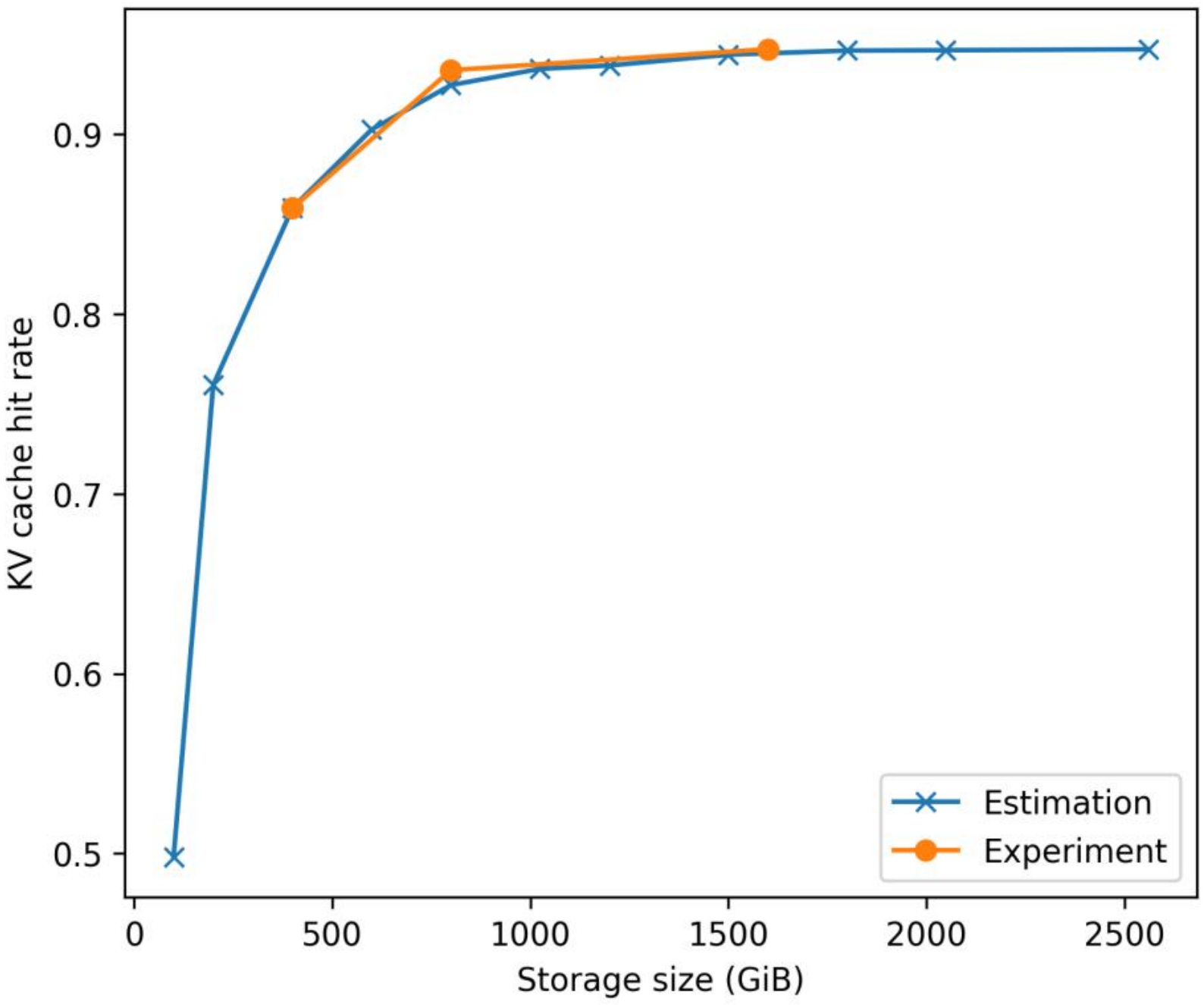

Figure 3. KV cache hit rate as a function of storage capacity. Estimation shows KVSET's predictions, whereas Experiment reports measurements from deployed caches. We measured only selected capacities because each deployment is time- and resource-intensive.

The cache hit rate increases rapidly as capacity grows in the low-capacity regime, but the marginal improvement diminishes substantially beyond a certain point. This behavior has two important implications. First, sufficient cache capacity is necessary to achieve a hit rate close to the maximum attainable for the workload. Second, as the hit rate approaches this maximum, increasingly large amounts of additional storage yield only marginal gains. Therefore, practical KV cache provisioning should carefully balance the storage cost against the throughput improvement enabled by a higher cache hit rate.

**Online KV Cache Capacity Analysis**

KVSET integrates with online serving systems to estimate KV cache capacity requirements without simulating many candidate cache sizes independently. We evaluate its online analysis capability using a production trace collected from an internal coding-agent service. Specifically, we analyze a sequence of 24,000 requests and track how the required KV cache capacity evolves as more requests are observed, as shown in Figure 4. The cumulative capacity represents the amount of storage required to retain the KV cache states generated by all historical requests. This cumulative capacity increases monotonically with the number of requests.

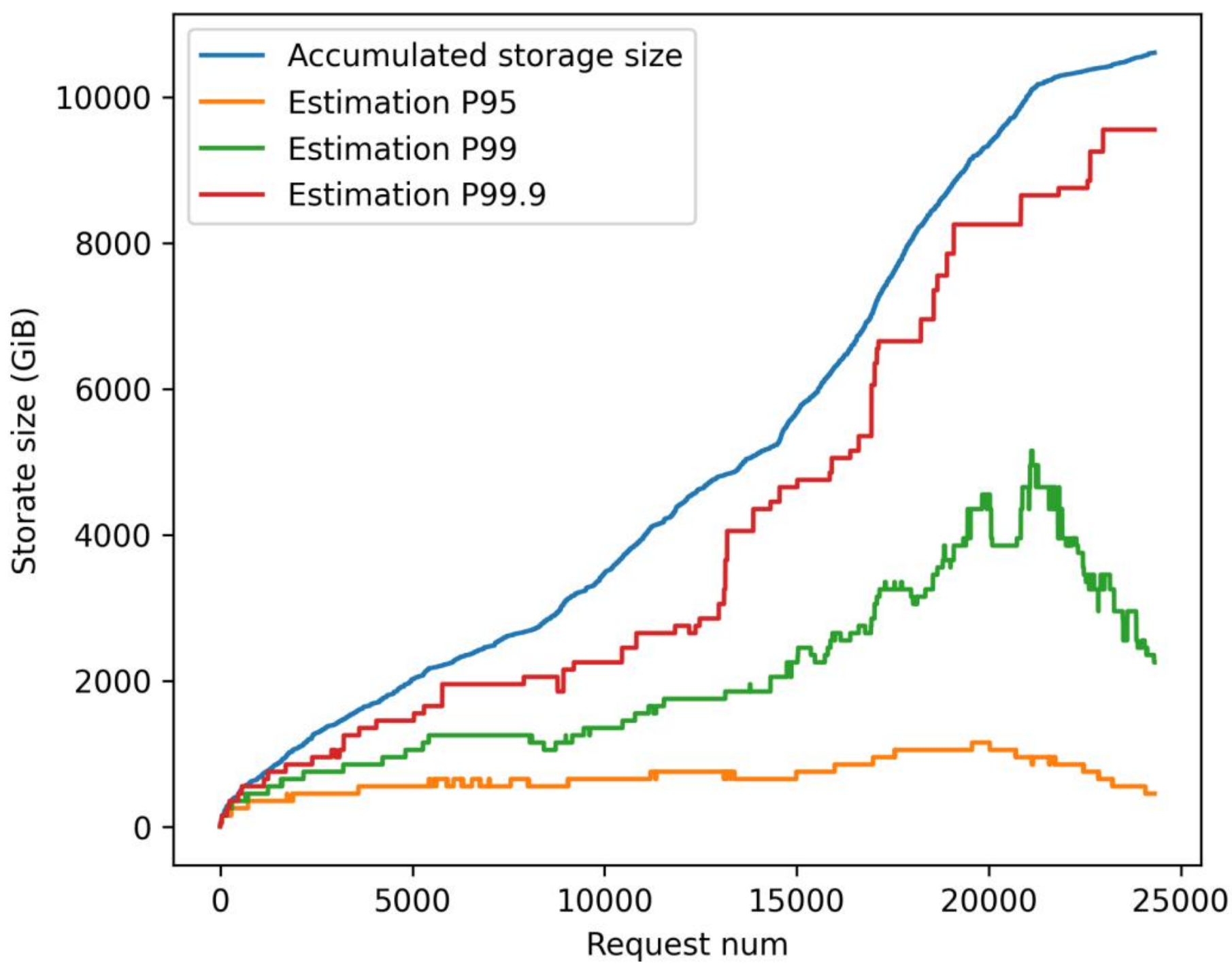

Figure 4. Accumulated and estimated KV cache storage capacity as the number of requests increases. Accumulated is the capacity required to retain all historical KV states, whereas Estimated is the capacity required for a specified fraction of requests to achieve their theoretical hit rates. The curves drop near 21,000 requests because the number of concurrently active users decreases.

The required cache capacity depends strongly on the coverage target: the fraction of requests whose theoretical hit rate the system aims to preserve. A request's theoretical hit rate is the rate achievable with unlimited KV cache capacity. At 99.9% coverage, the required cache capacity continues to grow with the request stream and shows no clear sign of convergence. In contrast, relaxing this requirement slightly leads to a substantial reduction in capacity. At 99% coverage, approximately 5 TiB of KV cache storage is sufficient for this workload. At 95% coverage, the required capacity decreases to approximately 1 TiB.

These results provide practical guidance for provisioning hierarchical KV cache storage. For this workload, consider a two-tier hierarchy of CPU memory and disk. If the CPU-memory and disk tiers target 95% and 99% request coverage, respectively, our analysis suggests provisioning approximately 1 TiB of CPU memory and 5 TiB of disk capacity for KV cache storage.

The appropriate coverage threshold ultimately depends on the trade-off between the performance benefit of improving the KV cache hit rate and the cost of provisioning additional storage. For example, a 5 TiB cache could fit within the aggregate CPU memory and local disk capacity of a single GPU server. Supporting substantially larger capacities, however, may require additional storage resources or external storage infrastructure, which introduces additional cost. Online capacity analysis therefore allows practitioners to select a KV cache size based on an explicit performance-cost trade-off rather than over-provisioning storage for the entire request history.

## 4 Related Work

**Cache working-set and miss-ratio analysis.** Mattson et al. introduced stack-distance analysis for deriving hit and miss ratios across all capacities of an LRU cache from one reference stream[1, 8]. Subsequent work reduced the overhead of online analysis. Counter Stacks estimates miss-ratio curves with sublinear space[15], while Cuki estimates dynamic working-set size and miss-ratio curves for variable-size objects and uses them for adaptive capacity tuning[4]. KVSET draws on this line of work but targets exact, page-granular analysis of LLM prefix caches. It converts token-prefix reuse into a KV-page reference stream and reports the cache capacity required to satisfy request-level hit-rate targets.

**KV cache systems and workload studies.** VLLM[6], SGLang[17], CachedAttention[3], and Mooncake[10] demonstrate the value of reusing and extending KV states across requests and storage tiers. KV Cache in the Wild[13] characterizes production KV reuse and shows that cache capacity and eviction policy materially affect hit rate and serving performance. Public tools such as kvcache-simulator[19] replay traces across multiple memory budgets and eviction policies. These systems and studies focus primarily on serving architecture, workload characterization, policy design, or offline capacity sweeps. KVSET is complementary: it maintains a single reuse-distance state to estimate many LRU capacities as requests arrive, avoiding a separate cache instance for every candidate capacity.

**Agentic-workload KV management.** CacheWise[12] and KVFlow[9] address long-running coding-agent and multi-agent workloads by improving scheduling, eviction, or prefix reuse. Their goal is to improve runtime cache behavior and end-to-end performance. KVSET instead asks a provisioning question: how much KV storage is needed for a specified fraction of requests to retain a target fraction of their theoretically reusable prefix. Thus, KVSET can inform DRAM and SSD tier sizing and can be used alongside these runtime optimizations.

# 5 Conclusion

We presented KVSET, an online analyzer for estimating the working-set capacity of page-granular LRU KV caches. KVSET combines the Mattson stack algorithm with a Fenwick tree to compute stack distances efficiently, derives hit rates for multiple cache capacities from a single trace replay, and identify the minimum capacity based on maximum LRU depth of prefix pages that meets a target hit rate. Experiments on production coding-agent traces show that these estimates closely match measurements from deployed caches, demonstrating that accurate capacity analysis does not require deploying or independently simulating each candidate cache size.

The results also show why cache provisioning should be based on an explicit coverage target rather than the total volume of historical KV states. In the evaluated workload, approximately 1 TiB and 5 TiB of storage were sufficient to preserve the theoretical hit rates of 95% and 99% of requests, respectively, while the capacity required for 99.9% coverage did not converge within the observed trace. With support for online request processing and offline trace replay, KVSET provides a practical basis for sizing hierarchical cache tiers and balancing storage cost against prefix-cache performance.

KVSET currently estimates capacity only for the full-attention component of a model. For hybrid architectures that combine full attention with linear attention or sliding-window attention (SWA), we recommend estimating the KV cache requirement of the full-attention layers and reserving additional memory for Mamba or SWA states using a ratio tailored to the specific model and workload. KVSET currently applies only to caches that use LRU eviction, which is the policy adopted by the mainstream inference engines and KV cache storage systems. Extending this analysis beyond LRU eviction and full-attention KV states remains future work.